\documentclass[12pt]{amsart}
\usepackage{amsmath}
\usepackage{latexsym}
\usepackage{amsfonts}
\usepackage{amssymb}

\newtheorem{proposition}{Proposition}

\newtheorem{corollary}{Corollary}
\theoremstyle{definition}

\newtheorem{example}{Example}

\newcommand{\mh}{\mathcal{H}}

\newcommand{\lh}{L(\mh)}

\newcommand{\R}{\mathbb R}
\newcommand{\C}{\mathbb C}
\newcommand{\Z}{\mathbb Z}
\newcommand{\T}{\mathbb T}
\newcommand{\D}{\mathbb D}
\newcommand{\N}{\mathbb N}

\newcommand{\br}{\mathcal B(\mathbb R)}
\newcommand{\bi}{\mathcal B([0,2\pi))}

\newcommand{\ip}[2]{\left\langle\,#1\,|\,#2\,\right\rangle}
\newcommand{\ket}[1]{\mid#1\rangle}
\newcommand{\kb}[2]{|#1\,\rangle\langle\,#2|}
\newcommand{\no}[1]{\parallel#1\parallel}

\def\tx{\textstyle}

\def\ov#1\overline{{#1}}

\begin{document}
\title[]{
Noise sequences of infinite matrices and their applications to the
characterization of the canonical phase and box localization observables
}
\author[Lahti]{Pekka Lahti}
\address{Pekka Lahti, Department of Physics, University of Turku, FIN-20014 Turku, Finland}
\email{pekka.lahti@utu.fi}
\author[Maczynski]{Maciej J. Maczynski}
\address{Faculty of Mathematics and Information Scienc, 
Warsaw University of Technology,
PL-00-661 Warszawa, Poland} 
\email{mamacz@mini.pw.edu.pl}
\author[Scheffold]{Egon Scheffold}
\address{Fachbereich Mathematik, Technische Universität Darmstadt, Darmstadt, Germany} 
\email{scheffold@mathematik.tu-darmstadt.de}
\author[Ylinen]{Kari Ylinen}
\address{Department of Physics, 
University of Turku,
FIN-20014 Turku, Finland}
\email{kari.ylinen@utu.fi}
\maketitle{}

\begin{abstract} 
Noise sequences of infinite matrices associated with covariant phase 
and  box localization observables are defined and determined. 
The canonical  observables are characterized within
the relevant classes of observables as those with   asymptotically minimal or minimal noise, i.e., the noise tending to 0 
or having the value 0.
\end{abstract}

\section{Introduction}
In this paper we study infinite matrices whose entries are complex numbers of absolute value at most 1 and which are indexed by
the sets $\N^2$ and $\Z^2$. In other words, we consider functions
$A:\N\times\N\to\D$ and
$B:\Z\times\Z\to\D$, where $\D=\{z\in\C\,|\, |z|\leq 1\}$. 
The sets $\mathcal F_{\D}^{\N}$ and $\mathcal F_{\D}^{\Z}$ of such functions are convex in a natural way and their extremal elements
are the torus valued functions
$A:\N\times\N\to\T$ and
$B:\Z\times\Z\to\T$,
respectively. 
For each matrix $A\in\mathcal F_{\D}^{\N}$, resp.  $B\in\mathcal F_{\D}^{\Z}$, we define a probabilistic concept called 
{\em noise sequence} (Sect.~\ref{nsn}, resp. Sect.~\ref{nsz}) which is used to characterize the extremal set 
$\mathcal F_{\T}^{\N}$,
resp. $\mathcal F_{\T}^{\Z}$ (Propositions~\ref{nnsn} and \ref{nnsz}, respectively). 

A function $A:\N\times\N\to\D$ is positive semidefinite if for each sequence 
$(c_n)_{n\in\N}\subset\C$, for which $c_n\ne 0$
for only finitely many $n\in\N$, $\sum_{n,m}\overline{c_n}A(n,m)c_m\geq 0$. Such a function  is normalized if $A(n,n)=1$ for all $n\in\N$.
Positive semidefinite normalized matrices $A\in\mathcal F_{\D}^{\N}$ have an important application in quantum mechanics as they
characterize the  $\N$-covariant semispectral measures, known also as the phase shift covariant phase observables.
Among them are the so-called canonical phase observable, associated with the constant one function, 
and its
unitary equivalents (in the sense of covariance systems) which are associated with the $\T$-valued functions. 
Proposition~\ref{nnsn} of Sec.~\ref{nsn} leads in Corollary~\ref{phasecor} to  a
characterization of the canonical phase observable (and its unitary equivalents) in terms of the noise sequence of
the associated matrix. 
This application will be discussed in Section~\ref{phase}.
Similarly, in Section~\ref{localization} we recall that the normalized positive semidefinite matrices $B:\Z\times\Z\to\D$ 
correspond bijectively to the  $\Z$-covariant semispectral measures,
known also as the translation (mod $2\pi$) covariant localization observables of a quantum object confined to move in a
one-dimensional box (of length $2\pi$).
Proposition~\ref{nnsz} of Section~\ref{nsz} serves to characterize those $\Z$-covariant semispectral measures which are projection valued,
and thus unitarily equivalent to the canonical spectral measure (in the sense of a covariance system). They are exactly those whose defining matrices
are noiseless (Corollary~\ref{boxcor}).

The structure of the $\N$ and the $\Z$ -covariant semispectral measures involves the Schur product of infinite matrices. 
In the final section of the paper we draw attention to that connection.

\section{The noise sequence of a matrix $A:\N\times\N\to\D$}\label{nsn}

To define the noise sequence of a matrix  we define first 
the discrete probability measures
$p_n:2^{\N}\to[0,1]$,  $n\in\N$, by the formula
\begin{equation}
p_n(\{k\})= 
\left\{\begin{array}{ll}
{3\over {\pi^2(k-n)^2}}, & \textrm{for $ k\ne n $},\\
{}\\
1-\sum_{l=0\atop l\ne n}^{\infty} {3\over {\pi^2 (l-n)^2}},& 
\textrm{for  
$k=n $}.
\end{array}
\right. 
\end{equation}
Since
$\sum_{l=0\atop l\ne n}^{\infty} {3\over {\pi^2(l-n)^2}} \leq 1$, we have
$\sum_{k=0}^\infty p_n(\{k\})=1$,
showing that $p_n$ is, indeed,  a well-defined probability.

Consider a matrix  $A\in\mathcal F^{\N}_{\D}$.
For each $n\in\N$,  define a random variable $X_n^A:\N\to\R$, $X_n^A(k)=a_{nk}^A$,
where
$$
a_{nk}^A= 
\left\{\begin{array}{ll} 
{{\pi}\over {\sqrt{3}}} |A(n,k)|,&\textrm{for $k\not= n$},\\
0, &\textrm{for $k=n$}.
\end{array}
\right. 
$$

The $l$-th moment of the random variable $X_n^A$  is
\begin{eqnarray*}
M^{(l)} (X_n^A)
&=& \sum_{k=0}^\infty X_n(k)^l p_n(\{k\})\\
 &=& {{\pi^{l-2}}\over 3^{\frac l2-1}}
 \sum_{k=0\atop k\ne 0}^{\infty} {{|A(n,k)|^l}\over {(k-n)^2}}, \ l\geq 1. 
\end{eqnarray*}

If $A$ is  the constant function  $A^{1}:\N\times\N\to\{1\}$, we write e.g. $X^1_n$ instead of $X^{A^1}_n$. For that we get
\begin{equation}
M^{(l)} (X^{1}_n)={{\pi^{l-2}}\over 3^{\frac l2-1}}
\sum_{k=0\atop k\ne n}^{\infty} {1\over {(k-n)^2}},
\end{equation}
and it is easy to see that
\begin{equation}
\lim_{n\to\infty} M^{(l)} (X_n^{1})
={{\pi^{l-2}}\over 3^{\frac l2-1}}
\left({{\pi^2}\over 6}+ {{\pi^2}\over 6}\right) 
=
\bigg({\pi\over{\sqrt{3}}}\bigg)^l.
\end{equation}
Clearly, if $A:\N\times\N\to\mathbb D$
takes  values in $\T$,
then $ M^{(l)} (X_n^{A})= M^{(l)} (X_n^{1})$ for all $n$ and $l$.

For any  function  $A:\N\times\N\to\mathbb D$ 
we  define, for each $n\in\N$, the numbers   
\begin{equation}\label{snal}
s_n^A (l)=\lim_{n\to\infty} M^{(l)} (X_n^{1})-M^{(l)}(X^A_n), \ l=0,1,2,\ldots,
\end{equation}
and we call the limiting sequence $(\lim_n s_n^A (l))_l$ the {\em asymptotic noise sequence} of $A$.

The numbers $s_n^A (l)$ 
are easily determined, and one gets
\begin{equation*}
s_n^A (l)= 
\bigg({\pi\over{\sqrt{3}}}\bigg)^l \bigg(1-{3\over {\pi^2}} 
\sum_{\tx{k=0\atop k\not= n}}^{\infty} {{|A(n,k)|^l}\over {(k-n)^2}}\bigg ) \geq 0.
\end{equation*}

We have following proposition. 

\begin{proposition}\label{nnsn}
 For any  function $A:\N\times\N\to\mathbb D$, 
if  $\lim_n s_n^A (l)=0$ for some $l=1,2,\ldots$, then $|A(n,m)|=1$ for all $n,m\in\N, n\ne m$.
Conversely, if $A$ takes values in $\T$, 
then $\lim_n s_n^A (l)=0$
for any $l=0,1,2,\ldots$.
\end{proposition}

\begin{proof}
Let us assume that $\lim_{n\to\infty} s_n^{A} (l)=0$ for some $l=1,2,\ldots.$ This implies that
\begin{equation}\label{apu6}
 \lim_{n\to\infty}
\sum_{\tx{k=0\atop k\not= n}}^{\infty} {{|A(n,k)|^l}\over {(k-n)^2}}={\frac{\pi^2}3}.
\end{equation}
We will show that then for all $n,k\in\N$, $n\ne k$, $|A(n,k)|=1$.
Assume that this is not the case, i.e., there are $n_o,k_o, n_o\ne k_o$, such that $|A_{n_ok_o}|<1$.
Then $|A_{n_ok_o}|^l=1-\epsilon$ for some $\epsilon>0$.
We infer that
$$
\lim_{n\to\infty}
\sum_{\tx{k=0\atop k\not= n}}^{\infty} {{|A(n,k)|^l}\over {(k-n)^2}}
\ <\ 
\lim_{n\to\infty}
\sum_{\tx{k=0\atop k\not= n}}^{\infty} 
{1\over {(k-n)^2}}
- {\epsilon \over{(k_o-n_o)^2}}
={\frac{\pi^2}3 }- {\epsilon\over{(k_o-n_o)^2}},
$$
which is a contradiction with (\ref{apu6}).
Hence, for any $k,n\in\N$, $n\ne k$, $|A(n,k)|=1$.
Clearly, if  $|A(n,m)|=1$ for all $n,m\in\N$, then $s^A_n(l)= s_n^{1} (l)$
so that $\lim_n s^A_n(l)=0$ for any $l=1,2,\ldots$. 
\end{proof}

If the matrix $A\in\mathcal F_{\D}^{\N}$ satisfies the condition of the above proposition we say that  $A$ is
{\em asymptotically noiseless}. We may thus conclude that the extremal elements  $A\in\mathcal F_{\T}^{\N}$ are exactly the 
asymptotically noiseless matrices $A$ with $|A(n,n)|=1$.

\begin{example}
To illustrate the above results we consider here a special class of functions $A^{\xi}:\N\times\N\to\D$, $0\leq \xi\leq 1$,
the {\em chessboard matrices} \cite{JPvk2}, defined as
$$
A^{\xi}(n,k)= 
\left\{\begin{array}{ll} 
1,&\textrm{for $n+k$} \, \in 2\, \N,\\
\xi, &\textrm{\ otherwise}.
\end{array}
\right. 
$$
For them the numbers (\ref{snal}) can easily be computed.  
For simplicity we write, for instance,  $s^\xi$ instead of $s^{A^\xi}$, and we get for each $k$ and $l$,
\begin{eqnarray*}
s^{\xi}_{2k+1}(l)&=&
{{\pi^{l-2}}\over 3^{\frac l2-1}}\bigg(\frac{\pi^2}{3}-\beta_{2k}-\xi^l\alpha_{2k+1}-\xi^l\alpha_\infty-\beta_\infty\bigg),\\
s^{\xi}_{2k}(l)&=&
{{\pi^{l-2}}\over 3^{\frac l2-1}}\bigg(\frac{\pi^2}{3}-\beta_{2k}-\xi^l\alpha_{2k-1}-\xi^l\alpha_\infty-\beta_\infty\bigg),
\end{eqnarray*}
where
$$
\beta_{2k} = \sum_{m=1}^k\frac 1{(2m)^2},\quad
\alpha_{2k+1}=\sum_{m=0}^k\frac 1{(2m+1)^2},\quad
\alpha_\infty = \frac{\pi^2}{8},\quad \beta_{\infty}=\frac{\pi^2}{24}.
$$ 
This gives us $s^{\xi}_{2k}(l)-s^{\xi}_{2k+1}(l)=\xi^l\frac 1{(2k+1)^2}\geq 0$, and
 $s^{\xi}_{2k+1}(l)-s^{\xi}_{2k+2}(l)=\frac 1{(2k+1)^2} > 0$, so that we get
$$
s^{\xi}_{2k}(l)\geq s^{\xi}_{2k+1}(l) > s^{\xi}_{2k+2}(l).
$$
For $\xi =1$ we obtain 
$$
s^1_0(l)=\frac{\pi^2}6,\quad \lim_{k\to\infty}s^1_{2k+1}(l)=0,\quad  \lim_{k\to\infty}s^1_{2k}(l)=0.
$$
Therefore, for any chessboard matrix $A^\xi$, $0\leq \xi\leq 1$, the asymptotic noise sequence for the $l$-th moment
$$
s^{\xi}(l)=(s^{\xi}_0(l), s^{\xi}_1(l),s^{\xi}_2(l),\ldots,s^{\xi}_k(l),\ldots)
$$
is a decreasing sequence with respect to $k$. The asymptotic noise is minimal for $\xi=1$, 
and the noise element $s^{\xi}_k(l)$ tends to 0 when $k\to\infty$.
\end{example}

\section{Application to covariant phase observables}\label{phase}

Let $\mh$ be a complex (separable) Hilbert space, $\lh$ the set of bounded operators on $\mh$, and $\mathcal N= (\ket n)_{n\in\N}$
an orthonormal basis  of $\mh$ labelled by $\N$.
Let $N$ be the selfadjoint operator with the property $N\ket n = n\ket n$, $n\in\N$. We call it the number operator with respect to the basis $\mathcal N$. 
Consider the unitary representation
$x\mapsto U_x=e^{ixN}$ of the real line $\R$. 
Let  $E:\bi\to\lh$ be a semispectral measure (a normalized positive operator measure)
defined on the Borel subsets of the interval $[0,2\pi)$. We say that $E$ is covariant under the shifts 
$U_x$  generated by the number operator $N$ 
if for any $x\in\R$ and $X\in\bi$,
\begin{equation}
U_xE(X)U_x^*= E(X\oplus x),
\end{equation}
where $X\oplus x=\{y\in[0,2\pi)\,|\, (y-x)({\rm mod} 2\pi)\in X\}$.
We call such semispectral measures  $\N$-covariant, or phase shift covariant phase observables, if we wish  to emphasize the physical
interpretation of these operator measures.

For each $X\in\bi$ define the function
$i_X:\N\times\N\to\D$ 
by the formula
$$
i_X(n,m)=\frac 1{2\pi}\int_Xe^{i(n-m)x}\,dx.
$$
 Let
$A:\N\times\N\to\D$ be  a normalized  positive semidefinite function.
The formula
\begin{equation}\label{phaseobservable}
E^A(X)=\sum_{n,m}A(n,m)i_X(n,m)\kb{n}{m}
\end{equation}
defines, in the weak sense, a positive operator bounded by the unit operator,
and the map $\br\ni X\mapsto E^A(X)\in\lh$ constitutes a semispectral measure.
A direct computation shows that it is $\N$-covariant. It is well known
that any $\N$-covariant semispectral measure $E$ is of the form (\ref{phaseobservable}) for a unique normalized
positive semidefinite matrix 
$A\in\mathcal F_{\D}^{\N}$ \cite{Holevo,JPvk1}. In the context of covariant phase observables the associated matrices are
also called the {\em phase matrices}.

The constant one function $A^1$ is a special case of a phase matrix. The $\N$-covariant
semispectral measure $E^1$, defined by it, is known as the canonical phase observable. 
The semispectral  measure $X\mapsto E^1(X)$ is unitarily equivalent to the
Toeplitz measure \cite{LY}. The function $A^1$ is a special case of the functions $A:\N\times\N\to\D$
taking values in the torus $\T$.
If $A:\N\times\N\to\mathbb T$ is normalized and positive semidefinite, then it is necessarily of the form 
$A(n,m)=e^{i(\nu_n-\nu_m)}$, $n,m\in\N$,  for some $\nu_n,\nu_m\in\R$.
This is exactly the case where the phase observable $X\mapsto E^A(X)$ is unitarily equivalent to the canonical
phase observable $X\mapsto E^1(X)$ (in the sense of a covariance system) \cite{JPvk2}.
 In that case we write $E^A\simeq E^1$.

Let $E^A$ be a phase shift covariant phase observable. All of  its (weakly defined) moment operators
$E^A[k]=\int_0^{2\pi}x^k\,dE^A(x)$, $k\in\N$, are bounded selfadjoint operators, and their explicit form can
easily be computed, see, for instance, \cite{JPvk2}. 
The difference of the second moment operator and the square of the first moment operator,
$S^A_E=E^A[2]-E^A[1]^2$, is positive  and it is called the {\em noise operator} of the phase observable $E^A$.
A direct computation shows that
the number $s_n^A(2)$ is the expectation value of the noise
operator of  
$E^A$ in the number state $\ket n$, that is, $s_n^A(2)=\ip{n}{S^A_E|n}$.
In the context of the phase observables the limit $\lim_{n\to\infty}\ip{n}{S^A_E|n}$ is the high-energy
limit of the noise of the phase observable.
The canonical phase and its unitary equivalents are thus distinguished as those phase observables which have {\em asymptotically
minimal noise}, tending to 0. We emphasize that the noise values $s_n^A(2)$ are always strictly positive and noise operator $S^A_E$ is
never zero, since no $E^A$ is projection valued.

We conclude this section with the following corollary of Proposition~\ref{nnsn} to  the theory of phase shift covariant phase observables.

\begin{corollary}\label{phasecor}
For a phase shift covariant phase observable $E^A$ 
the following conditions are equivalent:
\begin{itemize}
\item[(a)] $E^A$ has asymptotically minimal noise;
\item[(b)] $A$ is asymptotically noiseless;
\item[(c)] $A$ is $\T$-valued;
\item[(d)] $E^A\simeq E^1$.
\end{itemize}
\end{corollary}

\section{The noise sequence of functions $B:\Z\times\Z\to\D$}\label{nsz}

We consider the class of infinite matrices of the form
$B:\Z\times\Z\to\D$.
Analogously with Section~\ref{nsn}, we define, for each $n\in\Z$, 
the probability measure $p_n:2^\Z\to[0,1]$, with 
\begin{equation}\label{rv2}
p_n(\{k\})= 
\left\{\begin{array}{ll}
{3\over {\pi^2(k-n)^2}} &\textrm{for $k\not= n$}, \\
0& \textrm{for $k= n$},
\end{array}
\right.
\end{equation}
and
a  random variable $\Z\ni k\mapsto X_n^B(k)\in\R$,
with
$$
b_{nk}^B= 
\left\{\begin{array}{ll} 
{{\pi}\over {\sqrt{3}}} |B(n,k)|&\textrm{for $k\not= n$},\\
0 &\textrm{for $k= n$}.
\end{array}
\right. 
$$
Since 
$\sum_{k, k\not= n} {1\over {(k-n)^2}}=  {{\pi^2}\over{3}}$
for all $n\in\Z$, we have
$\sum_{k}p_n(\{k\})=1$,
showing that $p_n$  is a well-defined  probability measure.

We compute the $l$-th moment of the random variable $X_n^B$:
\begin{eqnarray*}
M^{(l)} (X_n^B)& =& \sum_{k=-\infty}^{\infty} X_n(k)^l p_n(\{k\}) \\
&=& \bigg({{\pi}\over {{\sqrt{3}}}}\bigg)^l
 \sum_{\tx{k=-\infty\atop k\not= n}}^\infty 
|B(n,k)|^l {3\over {\pi^2}} {1\over {(k-n)^2}}\\
& = &
\bigg({{\pi}\over {{\sqrt{3}}}}\bigg)^l{3\over {\pi^2}}
\sum_{\tx{k=-\infty\atop k\not= n}}^\infty 
{{|B(n,k)|^l}\over {(k-n)^2}}. 
\end{eqnarray*}
For $B= B^1$ this gives $M^{(l)} (X_n^1)= ({{\pi}\over {{\sqrt{3}}}})^l$ for all $n\in\Z$.

For any  matrix $B:\Z\times\Z\to\D$ and for each $n\in {\Z}$ we define
\begin{equation}
s_n^B(l) = M^{(l)} (X^{1})-M^{(l)} (X_n^B),
\end{equation}
and we call the sequence $(s_n^B(l))_{l\in\N}$ the {\em $n$-th noise sequence} of $B$.

We have
\begin{equation}\label{15}
s_n^B(l) =
\bigg({{\pi}\over {{\sqrt{3}}}}\bigg)^l
\bigg(1-\frac{3}{\pi^2}
\sum_{\tx{k=-\infty\atop k\not= n}}^{\infty} 
{{|B(n,k)|^l}\over {(k-n)^2}}\bigg). 
\end{equation}
In particular,  
$$
s_n^B(2) ={{\pi^2}\over{3}}
-\sum_{\tx{k=-\infty\atop k\ne n}}^{\infty} 
{{|B(n,k)|^2}\over {(k-n)^2}}. 
$$

\begin{proposition}\label{nnsz}
Assume that $B:\Z\times\Z\to\D$
is such that for some $l_o=1,2,\ldots$,
\begin{equation}\label{19}
s^B_{n}(l_o)=0,\ \ \ {\rm for\ all\ }n\in\Z.
\end{equation}
 Then $|B(n,k)|=1$ for all $n,k\in\Z,  n\ne k$, and
$s^A_{n}(l)=0$
for all $n\in\Z$ and $l=0,1,2\ldots$.
\end{proposition}

\begin{proof}
Assume that (\ref{19}) holds for some $l_o$ and for all $n$. Then by (\ref{15})  we have
\begin{equation}\label{20}
\sum_{\tx{k=-\infty\atop k\not= {n}}}^{\infty}{{|B(n,k)|^{l_o}}\over {(k-n)^2}}={\pi^2\over{3}}.
\end{equation}
Since $\sum_{k,k\ne n}  {{1}\over {(k-n)^2}}={3\over{\pi^2}}$,
analogously as in the  proof of Proposition~\ref{nnsn} we show that Equation  (\ref{20})
implies that 
for all $n,k\in\Z$,  $n\ne k$, $|B(n,k)|^{l_o}=1$, which gives the claim.
\end{proof}

If the matrix $B\in\mathcal F^{\Z}_{\D}$ fulfills the condition of Proposition~\ref{nnsz}, 
 then all the noise sequences of $B$ are 0-sequences. In that case we say that $B$ is {\em noiseless}.

\begin{example}
As an illustration we consider 
again the chessboard matrices, that is, the functions
$B^{\xi}:\Z\times\Z\to\D$, $0\leq \xi\leq 1$,
defined as
$$
B^{\xi}(n,k)= 
\left\{\begin{array}{ll} 
\xi,&\textrm{for $n+k$} \, \in 2\,\Z,\\
1, &\textrm{\ otherwise}.
\end{array}
\right. 
$$
The numbers (\ref{15}) can again explicitely be computed, and we get for each $n\in\Z$ and $l\in\N$:
$$
s^{\xi}_{n}(l) =
{{\pi^{l-2}}\over 3^{\frac l2-1}}\bigg(1-\xi^l\bigg)\,\frac{\pi^2}4.
$$
The noise elements $s^{\xi}_{n}(l) $ do not depend on $n$. In particular,
$$
s^{\xi}_{n}(2) =
\bigg(1-\xi^2\bigg)\,\frac{\pi^2}4.
$$
For any $l\geq 1$,
$s^{\xi}_{n}(l) =0$ if and only if $\xi=1$.
\end{example}

\section{Application to covariant localizations in a box}\label{localization}

Let $\mathcal Z= (\ket k)_{k\in\Z}$ be 
an orthonormal basis  of $\mh$ labelled by the integers $\Z$, 
$Z$ the associated selfadjoint operator ($Z\ket k = k\ket k$),
and $V_x=e^{ixZ}, x\in\R$. 
Extending the terminology of Section~\ref{phase}
we say that a semispectral measure 
$E:\bi\to\lh$ is  $\Z$-covariant if it satisfies the condition
\begin{equation}\label{boxcovariance}
V_xE(X)V_x^*= E(X\oplus x)
\end{equation}
 for all  $X\in\bi$, $x\in\R$.
In emphasizing the physical meaning of these  measures we call them (translation (mod $2\pi$) covariant) box localization observables.

For each $X\in\bi$ define the function
$i_X:\Z\times\Z\to\D$ as
$$
i_X(k,l)=\frac 1{2\pi}\int_Xe^{i(k-l)x}\,dx.
$$
If
$B:\Z\times\Z\to\D$ is  normalized  and positive semidefinite, then
the formula
\begin{equation}\label{boxobservable}
E^B(X)=\sum_{k,l}B(k,l)i_X(k,l)\kb{k}{l}, \ \ X\in\bi.
\end{equation}
defines, in the weak sense, a $\Z$-covariant semispectral measure. Conversely,
any $\Z$-covariant semispectral measure $E$ is of the form (\ref{boxobservable}) for a unique normalized
positive semidefinite matrix 
$B\in\mathcal F_{\D}^{\Z}$ \cite{Holevo,genova}. We call such a $B$ the {\em structure matrix} of $E$, and write
$E^B$ to indicate the one-to-one onto correspondence between the solutions of Eq. (\ref{boxcovariance}) and the normalized
positive semidefinite matrices $B\in\mathcal F_{\D}^{\Z}$.

The semispectral  measure $E^1$ associated with the constant one function $B^1$ is   the canonical spectral measure.
Moreover, an arbitrary $E^B$ is a spectral measure if and only if the structure matrix $B$ is $\T$-valued, and thus  of the form
$B(k,l)=e^{i(\nu_k-\nu_l)}$, $k,l\in\Z$  \cite{genova}. This is exactly the case
where the box observable $X\mapsto E^B(X)$ is unitarily equivalent to the canonical
spectral measure $X\mapsto E^1(X)$ (in the sense of a covariance system);
if this occurs we write $E^B\simeq E^1$ \cite{genova}.
Apart from the structural similarities between the  phase
and the box localization observables, it is, perhaps, worthwhile to recall that in the first case
there are no spectral measure solutions.

Consider a box localization
observable $E^B$. Its (weakly defined) moment operators $E^B[k]=\int_0^{2\pi}x^k\,dE^B(x)$ are bounded selfadjoint operators, and one may,
in particular, compute its noise operator $S_E^B= E^B[2]-E^B[1]^2$, which is a positive operator. Its expectation values
 with respect to the base states $|n\rangle$ are easily computed and one finds that they
coincide with the numbers $s_n^B(2)$ of the matrix $B$
\begin{equation}
\langle n|S_E^B|n\rangle = s_n^B(2),\ {\rm for\ all }\  n\in {\Z}.
\end{equation}
If  the  noise components $\langle n|S_E^B|n\rangle, n\in\Z$, are 0, 
then all the noise sequences $(s^B_n(l))_{l\geq 0}$, $n\in\Z$, are 0-sequences, and the
structure matrix $B$ is noiseless. 
But then also the observable $E^B$ is noiseless in the sense that its noise operator $S^B_E=0$;
this follows from the multiplicativity of a spectral measure. 
Therefore, we may conclude as follows.

\begin{corollary}\label{boxcor}
For a box localization observable $E^B$ the following conditions are equivalent:
\begin{itemize}
\item[(a)] $E^B$ is noiseless;
\item[(b)] $B$ is noiseless;
\item[(c)] $B$ is $\T$-valued;
\item[(d)] $E^B\simeq E^1$.
\end{itemize}
\end{corollary}

\section{A connection with the Schur product of matrices}

The structure of the $\N$- and $\Z$-covariant semispectral measures  involves the Schur product of the matrices
$A, i_X\in\mathcal F^{\N}_{\D,}$ and $B,i_X \in\mathcal F^{\Z}_{\D}$, respectively, see formulas (\ref{phaseobservable}) and (\ref{boxobservable}).
We consider here only the first case.
The Schur product of the infinite matrices $A$ and $i_X$ is  the pointwise product 
$$
(A\circ i_X)(n,m)= A(n,m)i_X(n,m),\ \ \ n,m\in\N.
$$
It is well known that the matrix $A\circ i_X$ defines, for all $X\in\mathcal B([0,2\pi))$, a positive operator
$E^A(X)$   bounded by the unit operator $I$  (on a separable Hilbert space $\mathcal H$)
if and only if $A$ is a normalized positive semidefinite matrix, which, in turn, is the case exactly when there is a
sequence of unit vectors $(\xi_n)_{n\in\N}$ such that $A(n,m)=\ip{\xi_n}{\xi_m}$ for all $n,m\in\N$, see, e.g. \cite{Holevo,JPvk1,genova}.
In particular, the matrix $i_X$, being the Schur product of $i_X$ with the constant one matrix $A^1$, defines an operator $0\leq E^1(X)\leq I$.

The above result can be seen as a special case of the theory of Schur multipliers. 
A matrix $A$ is called a Schur multiplier if for any matrix $B$,
which defines  a bounded operator on $l^2(\N)$, also the Schur product $A\circ B$ defines  a bounded operator on  $l^2(\N)$. 
A matrix $A$ is known to be a Schur multiplier if and only if there is a (complex separable) Hilbert space $\mathcal H$
with two bounded sequences of vectors $(\xi_n)_{n\in\N}$ and $(\eta_m)_{m\in\N}$ such that for all $n,m\in\N$, $A(n,m)=\ip{\xi_n}{\eta_m}$.
Moreover, a matrix $A$ defines a positive Schur multiplier (that is, $A\circ B$ is a bounded positive operator on $l^2(\N)$ whenever $B$ is a bounded positive operator on $\l^2(\N)$)
if and only if there is a Hilbert space $\mathcal H$ and a bounded sequence of vectors $(\xi_i)$ in $\mathcal H$ such that $A(n,m)=\ip{\xi_n}{\xi_m}$.
(See, e.g. Corollary 8.8. and Exercise 8.7. in \cite{Paulsen}.)

A matrix $A$, with $\sup_{n,m}|A(n,m)|<\infty$, does not necessarily define a Schur multiplier, see, e.g. Theorem 2.3 of \cite{Davidson}.
To give an explicit example, let $|A|= (|A(n,m)|)$ denote the modulus of $A$, and define $M_A$ to be the matrix $(c(n,m))$ such that $c(n,m)=0$, if
$A(n,m)=0$, and otherwise $c(n,m)=\overline{A(n,m)}/|A(n,m)|$. Then the matrix $M_A$ is bounded and $M_A\circ A=|A|$.
Consider now the $2^p$-dimensional Euclidean space $E_p$, $p=1,2,\ldots$. For each $p$ there exists a matrix $A_p$ with
$\no{A_p}=1$ and $\no{|A_p|}=\sqrt{2^p}$ \cite[Example 1, p. 231]{Schaefer}.  Now let $E$ be the Hilbert sum of the Hilbert spaces $E_p$,
$p=1,2,\ldots$, and let $A$ be the following operator on $E$ defined by $A(x_p)=(A_px_p)$ for all $(x_p)\in E$.
Since $\no{|A_p|}=\sqrt{2^p}\to\infty $  for $p\to\infty$, the matrix $A$ defines a bounded operator whereas its modulus $|A|$ does not.
When the Hilbert space $E$ is identified with $l^2(\N)$,  the matrix $M_A$ is  not a Schur multiplier.

For each $X\in\mathcal B([0,2\pi))$, the matrix $i_X$ defines a bounded positive operator $E^1(X)$. 
However, itsmodulus $|i_X|$ of $i_X$ does not necessarily define a bounded operator, showing that the corresponding $M_{i_X}$ is not
a Schur multiplier. 
To give an example we recall first some facts about the spectral radius and the $l^2$-norm
of a nonnegative, symmetric finite matrix.
Let  $B=(b_{ik})$ be an $n\times n$-matrix with  $b_{ik} \geq 0$ and  $b_{ik} = b_{ki}$. 
Let $r(B)$ denote the spectral radius of $B$, $s$ the smallest row sum and $t$ the greatest
row sum of $B$. Then
$s\leq r(B)\leq t$  \cite[Chap. I, Prop. 2.4]{Schaefer}.
Since the matrix $B$ is symmetric, we have  $\no{B} = r(B)$ and therefore
the inequality 
\begin{equation}\label{ineq}
 s\leq ||B|| \leq t.
\end{equation}
Choose now $X=[0,\pi]$. By calculation we get
$$
i_X(n,m)= 
\left\{\begin{array}{ll} 
\frac 12,&\textrm{\ for $n=m$},\\
0, &\textrm{\  for } 0\not= n-m  \textrm{\  even,}\\
\frac{i}{\pi (n-m)}, &\textrm{\  for } 0\not= n-m  \textrm{\  odd,}
\end{array}
\right.
$$

$$
M_{i_X}(n,m)= 
\left\{\begin{array}{ll} 
1,&\textrm{\ for $n=m$},\\
0, &\textrm{\  for } 0\not= |n-m|  \textrm{\  even,}\\
i &\textrm{\  for } 0< n-m  \textrm{\  odd,}\\
-i &\textrm{\  for } 0> n-m  \textrm{\  odd,}
\end{array}
\right.
$$
 and
 $$
 |i_X(n,m)|= 
\left\{\begin{array}{ll} 
\frac 12,&\textrm{\ for $n=m$},\\
0, &\textrm{\  for } 0\not= |n-m|  \textrm{\  even,}\\
\frac 1{\pi|n-m|} &\textrm{\  for } 0\not= |n-m|  \textrm{\  odd.}\\
\end{array}
\right.
$$
For $r\geq 1$,  let now be  $B_r = (|i_X(n,m)|)$, $0\leq n,m \leq r$,  and
$s_r$ the smallest row sum of $B_r$. Then $B_r$ is nonnegative and symmetric,
and $s_r$ is equal the sum of the first row of $B_r$. Therefore if  $r\geq 5$ and odd,
we have  $s_r  > (1+1/3+1/5+...+1/r)/\pi =: u_r$.
Now by (\ref{ineq}) we have for $r\geq 5$ and odd:
 $\no{B_r} \geq s_r > u_r$. Since $u_r\to\infty$ for $r\to\infty$, 
it follows that the norms of $B_r$ ($r$ natural number) are not
uniformly bounded. It is well known that then the matrix $|i_X|$ does not
define a bounded operator on $l^2(\N)$
\cite[p. 226]{Taylor}. 
Since $M_{i_X}\circ i_X=|i_X|$ for any $X$, we see that the matrix $M_{i_{ [0,\pi]}}$  is not a Schur multiplier.

\end{document}